\documentclass{PoS}
 
\newcommand{\Pom}{\mathbb{P}}

\newcommand{\Reg}{\mathbb{R}}

\title{Exclusive diffractive production of $\pi^+ \pi^-$ pairs 
within tensor pomeron approach}

\ShortTitle{Exclusive diffractive production of $\pi^+ \pi^-$ pairs within tensor pomeron approach} 

\author{\speaker{Piotr LEBIEDOWICZ}\\
Institute of Nuclear Physics, Polish Academy of Sciences, PL-31-342 Cracow, Poland\\
E-mail: \email{Piotr.Lebiedowicz@ifj.edu.pl}} 

\author{Otto NACHTMANN\\
Institut f\"ur Theoretische Physik, Universit\"at Heidelberg, Philosophenweg 16, D-69120 Heidelberg, Germany\\
E-mail: \email{O.Nachtmann@thphys.uni-heidelberg.de}}

\author{Antoni SZCZUREK\\
Institute of Nuclear Physics, Polish Academy of Sciences, PL-31-342 Cracow, Poland
\footnote{Also at University of Rzesz\'ow, PL-35-959 Rzesz{\'o}w, Poland.}\\
E-mail: \email{Antoni.Szczurek@ifj.edu.pl}}

\abstract{
We discuss exclusive central diffractive production of $\pi^{+} \pi^{-}$ in proton-(anti)proton collisions 
at high energies. Based on a tensor pomeron model 
we present results of the purely diffractive dipion continuum, 
the scalar $f_{0}(500)$, $f_{0}(980)$ and tensor $f_{2}(1270)$ resonances decaying into the $\pi^{+} \pi^{-}$ pairs as well as the photoproduction mechanism ($\rho^{0}$, Drell-S\"oding). 
We discuss how two pomerons couple to the tensor meson $f_{2}(1270)$ and the interference effects of resonance and dipion continuum contributions. The theoretical results are compared with existing STAR, CDF, and CMS experimental data. Predictions for planned or being carried out experiments (ALICE, ATLAS) are presented. 
We find that the relative contribution of resonant $f_2(1270)$ and dipion continuum strongly depend on the cut on proton transverse momenta (or four-momentum transfer squared $t_{1,2}$) which may explain some controversial observations made by different ISR experiments in the past.
The cuts may play then the role of a $\pi \pi$ resonance filter.
}

\FullConference{XXIV International Workshop on Deep-Inelastic Scattering and Related Subjects\\
          11-15 April, 2016\\
          DESY Hamburg, Germany} 
          
\begin{document}

\section{Introduction}

Recent theoretical \cite{Lebiedowicz:2014bea,Lebiedowicz:2016ioh,Lebiedowicz:2016zka} 
and experimental 
\cite{Adamczyk:2014ofa,Aaltonen:2015uva,Staszewski:2011bg,CMS:2015diy,CMS_talk} studies
have demonstrated that the exclusive dipion production
can be important in the context of resonance production, in particular, in searches for glueballs.
For a related work see \cite{Fiore:2015lnz}.
The experimental data on central exclusive $\pi^{+}\pi^{-}$ production
measured at the energies of the ISR, RHIC, Tevatron, and the LHC colliders 
all show visible structures in the $\pi^{+}\pi^{-}$ invariant mass.
Some time ago two of us have formulated a Regge-type model 
of the dipion continuum
for the exclusive reaction $pp \to pp \pi^{+}\pi^{-}$ with parameters 
fixed from phenomenological analysis of total and elastic $NN$ and $\pi N$ 
scattering \cite{Lebiedowicz:2009pj}.
The number of free model parameters is then limited to a parameter
of form factor describing off-shellness of the exchanged pion.
The model was extended to include rescattering corrections 
due to $pp$ nonperturbative interaction 
\cite{Lebiedowicz:2011nb,Lebiedowicz:2011tp}.
The largest uncertainties in the model are due
to the unknown off-shell pion form factor and the absorption effects
discussed recently in \cite{Lebiedowicz:2015eka}.
Such an approach gives correct order of magnitude cross sections,
however, does not include resonance contributions which interfere with the continuum contribution.

First calculations of central exclusive diffractive production of $\pi^{+} \pi^{-}$ continuum 
together with the dominant scalar $f_{0}(500)$, $f_{0}(980)$, 
and tensor $f_{2}(1270)$ resonances was performed in Ref.~\cite{Lebiedowicz:2016ioh}.
Here we based on the tensor-pomeron model 
formulated in \cite{Ewerz:2013kda}; see also \cite{Nachtmann:1991ua}.
In this model pomeron exchange is effectively treated as the
exchange of a rank-2 symmetric tensor.
In \cite{Ewerz:2016onn} we show that the tensor pomeron is 
consistent with the STAR data \cite{Adamczyk:2012kn}.
The corresponding couplings of the tensorial object to proton and pion were worked out.
In Ref.~\cite{Lebiedowicz:2013ika} the model was applied to 
the diffractive production of several scalar and pseudoscalar mesons in the reaction $p p \to p p M$.
We discussed there differences between results obtained with the "tensorial" and "vectorial" pomeron approaches.
In most cases one has to add coherently amplitudes for two pomeron-pomeron-meson couplings 
with different orbital angular momentum and spin of two "pomeron particles" 
\footnote{We wish to emphasize that the tensorial pomeron can, at least, equally well describe
the WA102 experimental data on the exclusive meson production as the less 
theoretically justified vectorial pomeron frequently used in the literature. 
The existing low-energy experimental data do not allow to
clearly distinguish between the two models as the presence of subleading
reggeon exchanges is at low energies very probable for many $p p \to p p M$ reactions.}.
In \cite{Bolz:2014mya} an extensive study of the photoproduction reaction
$\gamma p \to \pi^{+} \pi^{-} p$ was presented.
The resonant ($\rho^0 \to \pi^{+}\pi^{-}$) and non-resonant (Drell-S\"oding)
photon-pomeron/reggeon $\pi^{+} \pi^{-}$ production in $pp$ collisions
was studied in \cite{Lebiedowicz:2014bea}. 
Recently, in Ref.~\cite{Lebiedowicz:2016zka}, we analysed also
the exclusive diffractive production of $\pi^{+} \pi^{-}\pi^{+} \pi^{-}$
via the intermediate $\sigma \sigma$ and $\rho \rho$ states.

\section{Sketch of formalism}

\begin{figure} 
\centering
(a)\includegraphics[width=0.26\textwidth]{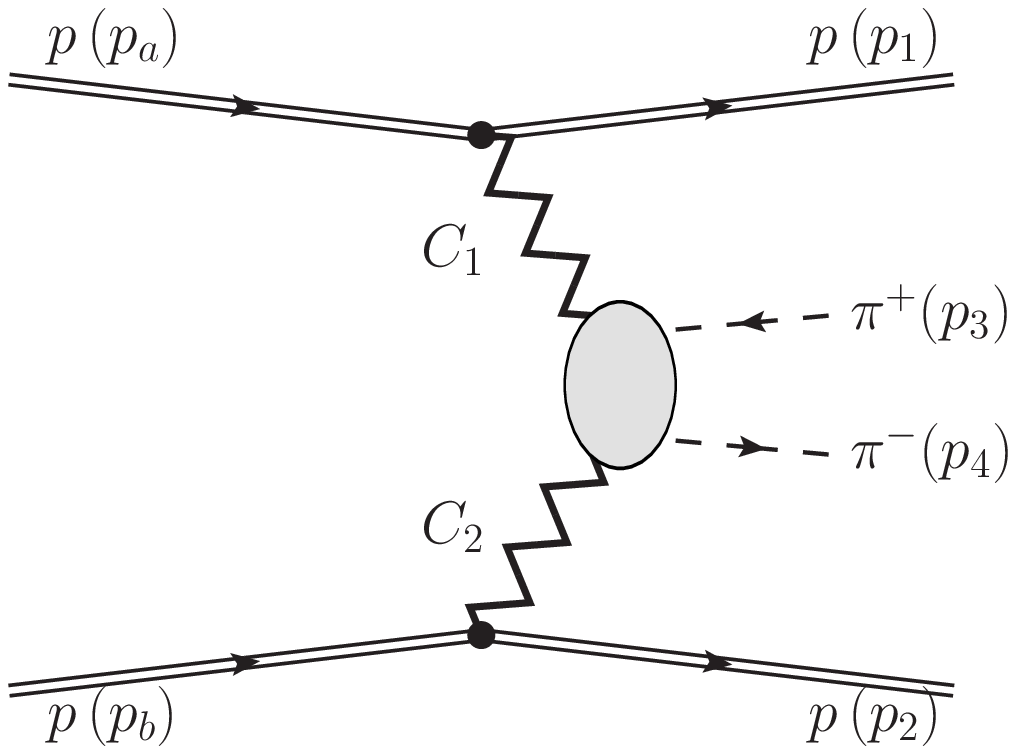}
(b)\includegraphics[width=0.32\textwidth]{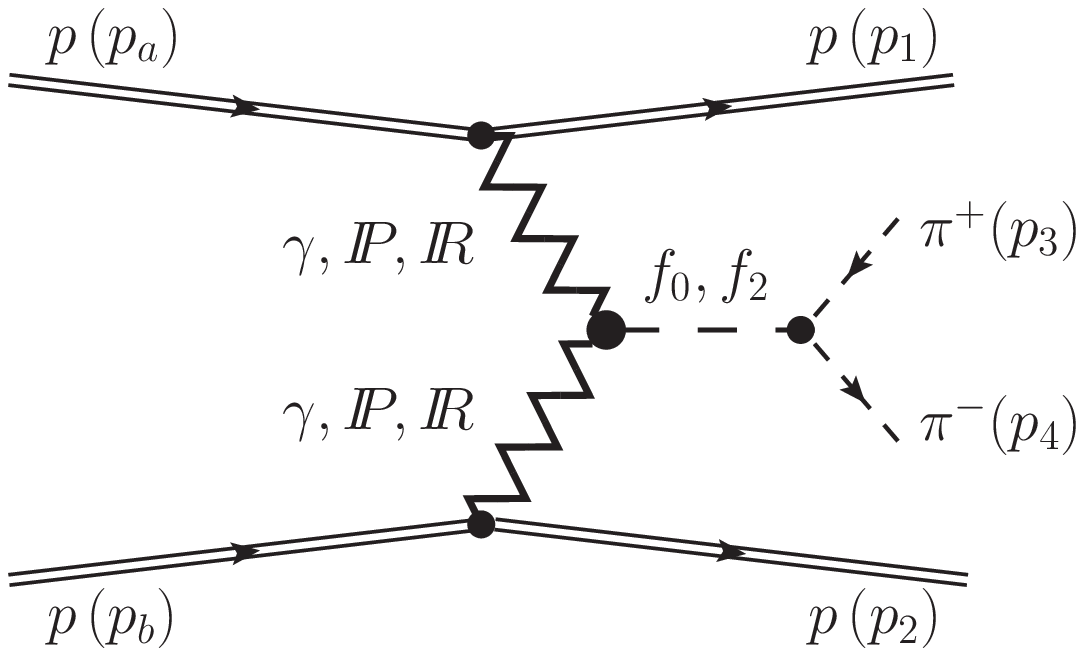}
(c)\includegraphics[width=0.28\textwidth]{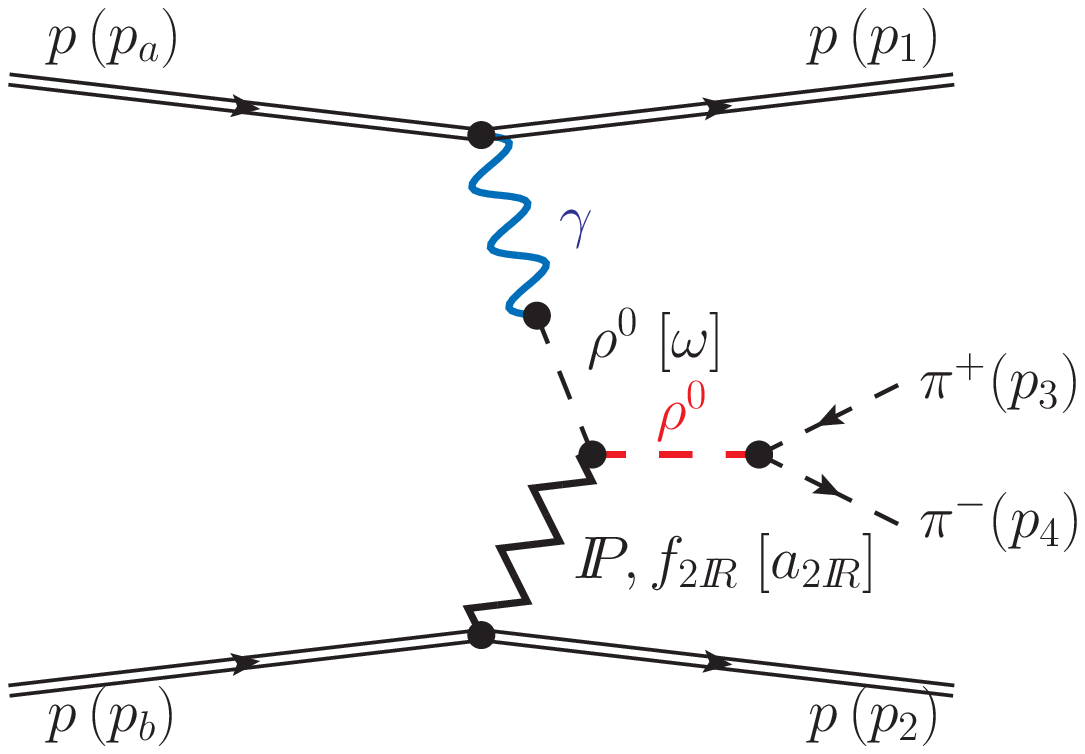}
\caption{
Diagram (a): Generic Born level diagram for central exclusive continuum $\pi^+ \pi^-$ production 
in proton-(anti)proton collisions. 
Here we labelled the exchange objects by their charge conjugation numbers
$C_{1}$, $C_{2} \in \lbrace +1, -1 \rbrace$.
Diagram (b): The double-pomeron/reggeon and photon mediated
central exclusive scalar and tensor resonances production 
and their subsequent decays into $\pi^+ \pi^-$.
Diagram (c): The $\rho(770) \to \pi^+ \pi^-$ photoproduction mechanism.
}
\label{fig:0}
\end{figure}

This amplitude for central exclusive $\pi^+ \pi^-$ production
is believed to be given by the ``fusion'' of two exchanged objects.
The Born level diagrams for the continuum and resonant production
are shown in Fig.~\ref{fig:0}.
The full amplitude of $\pi^{+} \pi^{-}$ production 
is a sum of continuum amplitude 
and the amplitudes with the $s$-channel resonances:
%
\begin{equation}
{\cal M}_{pp \to pp \pi^{+} \pi^{-}} =
{\cal M}^{\pi \pi{\rm-continuum}}_{pp \to pp \pi^{+} \pi^{-}} + 
{\cal M}^{\pi \pi{\rm-resonances}}_{pp \to pp \pi^{+} \pi^{-}}\,.
\label{amplitude_pomTpomT}
\end{equation}
The formulae for amplitudes
are presented and discussed in detail in Refs.~\cite{Lebiedowicz:2013ika,Lebiedowicz:2014bea,Lebiedowicz:2016ioh}.
For instance, the amplitude for the $\pi \pi$ production through 
the $s$-channel $f_{2}$-meson exchange 
can be written as
\begin{eqnarray}
&&{\cal M}^{(\Pom \Pom \to f_{2}\to \pi^{+}\pi^{-})}_{\lambda_{a} \lambda_{b} \to \lambda_{1} \lambda_{2} \pi^{+}\pi^{-}} 
=  (-i)\,
\bar{u}(p_{1}, \lambda_{1}) 
i\Gamma^{(\Pom pp)}_{\mu_{1} \nu_{1}}(p_{1},p_{a}) 
u(p_{a}, \lambda_{a})\;
i\Delta^{(\Pom)\, \mu_{1} \nu_{1}, \alpha_{1} \beta_{1}}(s_{1},t_{1}) \nonumber \\
&& \qquad \qquad \qquad \qquad \times 
i\Gamma^{(\Pom \Pom f_{2})}_{\alpha_{1} \beta_{1},\alpha_{2} \beta_{2}, \rho \sigma}(q_{1},q_{2}) \;
i\Delta^{(f_{2})\,\rho \sigma, \alpha \beta}(p_{34})\;
i\Gamma^{(f_{2} \pi \pi)}_{\alpha \beta}(p_{3},p_{4}) \nonumber \\
&& \qquad \qquad \qquad \qquad \times 
i\Delta^{(\Pom)\, \alpha_{2} \beta_{2}, \mu_{2} \nu_{2}}(s_{2},t_{2}) \;
\bar{u}(p_{2}, \lambda_{2}) 
i\Gamma^{(\Pom pp)}_{\mu_{2} \nu_{2}}(p_{2},p_{b}) 
u(p_{b}, \lambda_{b}) \,,
\label{amplitude_f2_pomTpomT}
\end{eqnarray}
where $t_{1} =(p_{1} - p_{a})^{2}$, $t_{2} =(p_{2} - p_{b})^{2}$, $s_{1} = (p_{a} + q_{2})^{2} = (p_{1} + p_{34})^{2}$,
$s_{2} = (p_{b} + q_{1})^{2} = (p_{2} + p_{34})^{2}$, and
$p_{34} = p_{3} + p_{4}$. $\Delta^{(\Pom)}$ and $\Gamma^{(\Pom pp)}$ 
denote the effective propagator and proton vertex function, respectively, for the tensorial pomeron.
For the explicit expressions, see Sec.~3 of \cite{Ewerz:2013kda}.
The pomeron-pomeron-$f_{2}$ vertex is the most complicated element of our amplitude (\ref{amplitude_f2_pomTpomT}).
In Ref.~\cite{Lebiedowicz:2016ioh} we have considered all possible tensorial structures 
for the $\Pom \Pom f_{2}$ coupling (see Appendix~A of \cite{Lebiedowicz:2016ioh})
\begin{eqnarray}
i\Gamma_{\mu \nu,\kappa \lambda,\rho \sigma}^{(\Pom \Pom f_{2})} (q_{1},q_{2}) =
\left( i\Gamma_{\mu \nu,\kappa \lambda,\rho \sigma}^{(\Pom \Pom f_{2})(1)} \mid_{bare}
+ \sum_{j=2}^{7}i\Gamma_{\mu \nu,\kappa \lambda,\rho \sigma}^{(\Pom \Pom f_{2})(j)}(q_{1},q_{2}) \mid_{bare} 
\right)
\tilde{F}^{(\Pom \Pom f_{2})}(q_{1}^{2},q_{2}^{2},p_{34}^{2}) \,.
\label{vertex_pompomT}
\end{eqnarray}
%
%
%
We take here the same form factor $\tilde{F}^{(\Pom \Pom f_{2})}$ for each vertex with index $j$ ($j = 1, ..., 7$).
Other details as form of form factors, the tensor-meson propagator $\Delta^{(f_{2})}$ and 
the $f_{2} \pi \pi$ vertex are given in Refs.~\cite{Ewerz:2013kda,Lebiedowicz:2016ioh}.
The production of the $f_{2}$ meson via $\Pom f_{2 \Reg}$, $f_{2 \Reg} \Pom$,
and $f_{2 \Reg} f_{2 \Reg}$ fusion can be treated 
in a completely analogous way to the $\Pom \Pom$ fusion.

We consider also the production of $\rho(770)$ resonance
produced by photon-pomeron/reggeon mechanism studied in detail in \cite{Lebiedowicz:2014bea}, 
see the panel (c) in Fig.~\ref{fig:0}.
In the amplitude for the $\gamma p \to \rho^{0} p$ subprocess
we included both pomeron and $f_{2 \Reg}$ exchanges.
The $\Pom \rho \rho$ vertex is given in \cite{Ewerz:2013kda} by formula (3.47).
The coupling parameters of tensor pomeron/reggeon exchanges was fixed based on
the HERA experimental data for the $\gamma p \to \rho^{0} p$ reaction.
In \cite{Lebiedowicz:2014bea} we showed that the resonant contribution
interfere with the non-resonant (Drell-S\"oding) $\pi^{+} \pi^{-}$ continuum
and produces a skewing of the $\rho(770)$-meson line shape.
Due to the photon propagators occurring in diagrams
we expect these processes to be most important when at least one of the protons
undergoes only a very small momentum transfer.
\section{Selected results}
\label{section:Results}

\begin{figure}[!ht]
\centering
\includegraphics[width=0.48\textwidth]{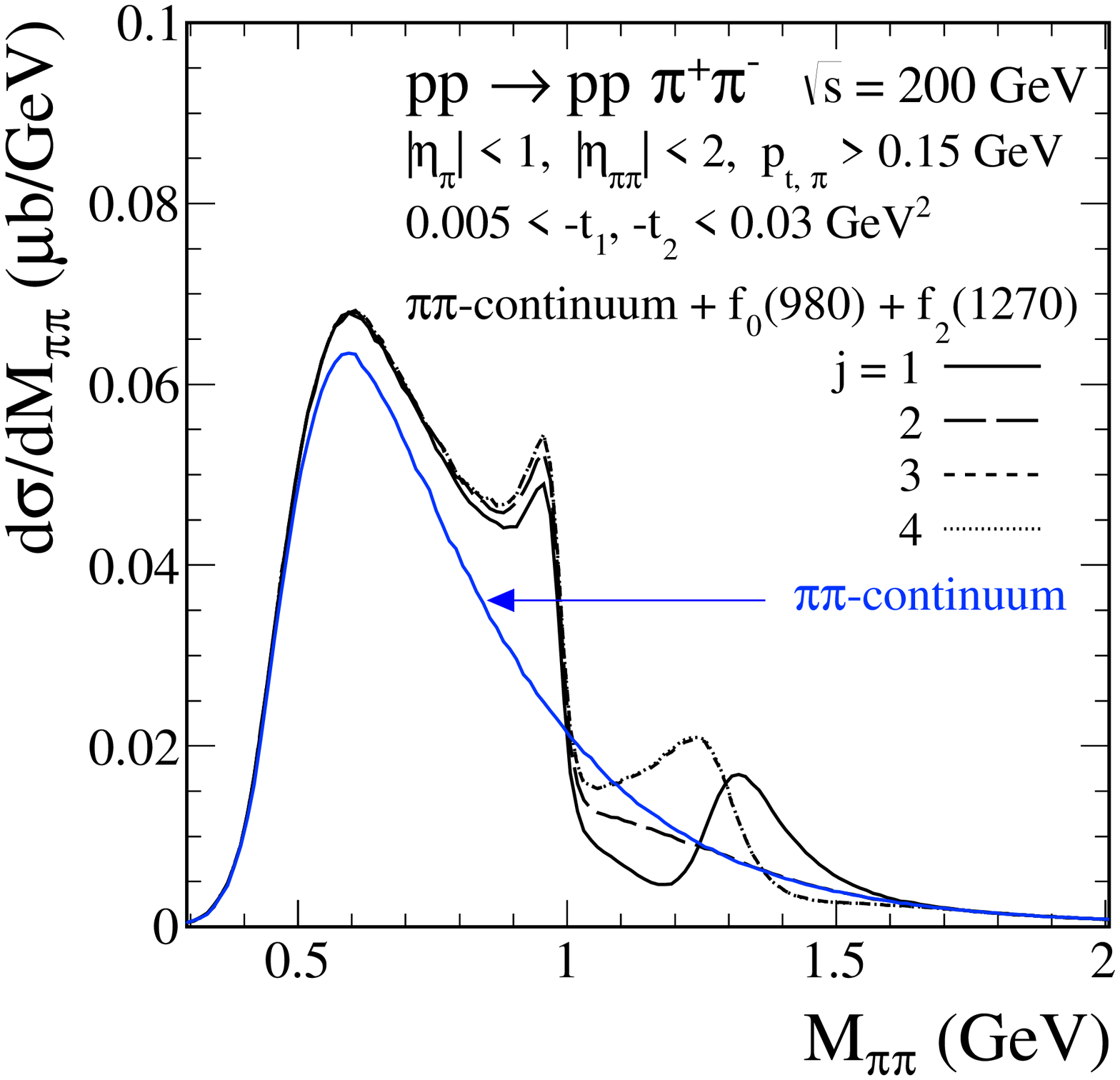}
\includegraphics[width=0.48\textwidth]{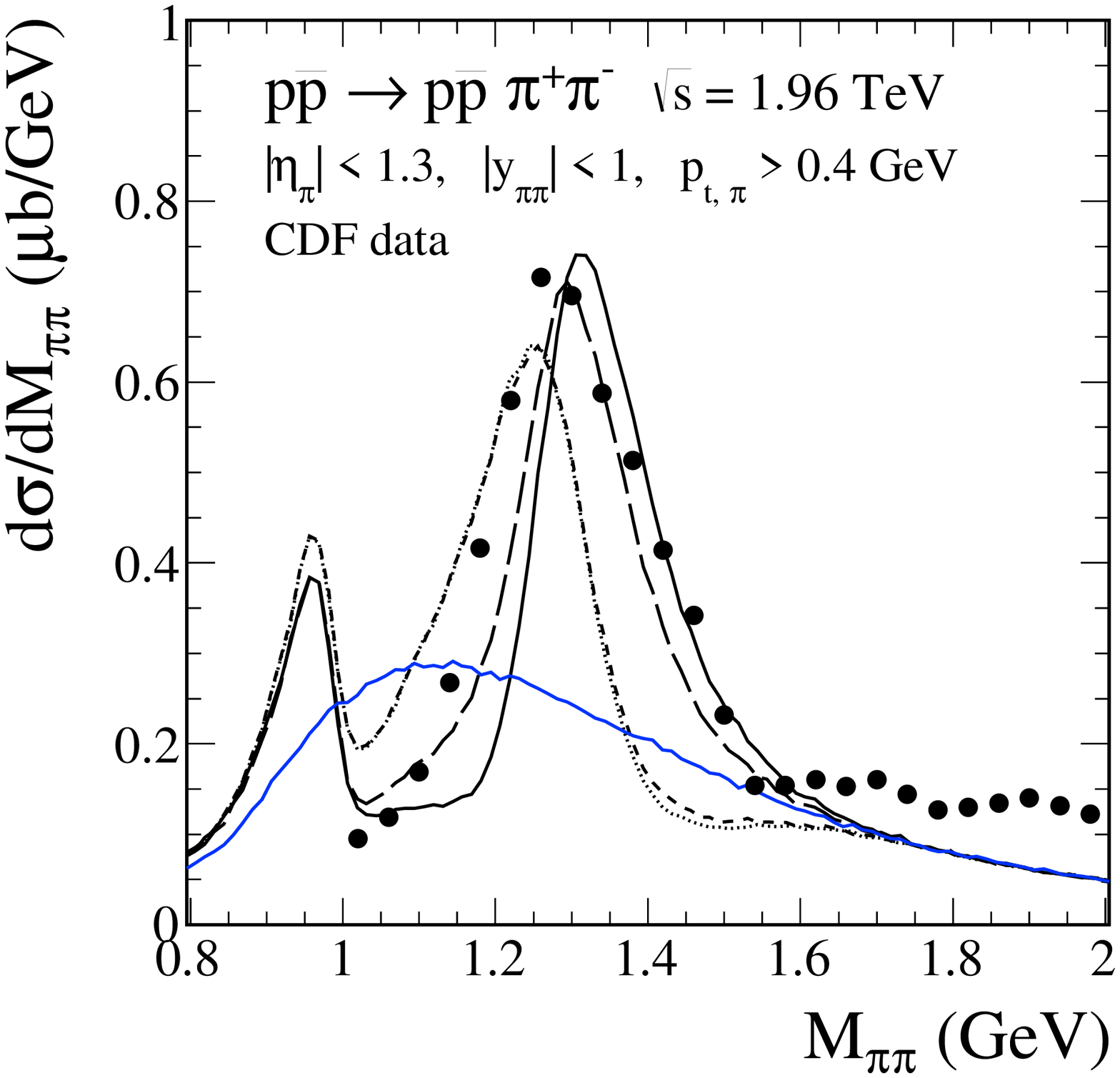}
  \caption{\label{fig:1}
  \small
Two-pion invariant mass distribution for the STAR \cite{Adamczyk:2014ofa}
and CDF \cite{Aaltonen:2015uva} kinematics.
The Born calculations for $\sqrt{s}=200$~GeV 
and $\sqrt{s}=1.96$~TeV were multiplied 
by the gap survival factors $\langle S^{2}\rangle = 0.2$
and $\langle S^{2}\rangle = 0.1$, respectively.
The blue solid lines represent the non-resonant continuum contribution only
while the black lines represent a coherent sum of non-resonant continuum, 
$f_{0}(980)$ and $f_{2}(1270)$ resonant terms. 
The individual contributions of different $\Pom \Pom f_{2}$ couplings ($j = 1, ..., 4$)
are compared with the CDF data \cite{Aaltonen:2015uva} (right panel).
}
\end{figure}

In our recent preliminary analysis \cite{Lebiedowicz:2016ioh}
we tried to understand whether one can approximately describe
the dipion invariant mass distribution observed by different experiments
assuming only one of seven possible $\Pom \Pom f_{2}$ tensorial couplings. 
The calculations were done at Born level
and the absorption corrections were taken into account
by multiplying the cross section
by a common factor $\langle S^{2}\rangle$ obtained from \cite{Lebiedowicz:2015eka}.
The two-pion continuum was fixed by choosing a form factor for the off-shell pion 
$\hat{F}_{\pi}(k^{2})=\frac{\Lambda^{2}_{off,M} - m_{\pi}^{2}}{\Lambda^{2}_{off,M} - k^{2}}$ and
$\Lambda_{off,M} = 0.7$~GeV.
As can be clearly seen from Fig.~\ref{fig:1} 
different couplings generate different interference patterns.
For detailed study of $f_{2}$ production see Ref.~\cite{Lebiedowicz:2016ioh}.
We found that the pattern of visible structures depends in addition
on the cuts used in a particular experiment
(usually the $t$ cuts are different for different experiments).
In Fig.~4 there we show that only in two cases ($j=2$ and 5)
the cross section $d\sigma/d|t|$ vanishes when $|t| \to 0$.
We can observe that the $j=2$ coupling gives results close to those observed by the CDF Collaboration \cite{Aaltonen:2015uva}.
In this preliminary study we did not try to fit the existing data \cite{Aaltonen:2015uva} 
by mixing different couplings because the CDF data are not fully exclusive
(the outgoing $p$ and $\bar{p}$ were not measured).

\begin{figure}[!ht]
\centering
\includegraphics[width=0.48\textwidth]{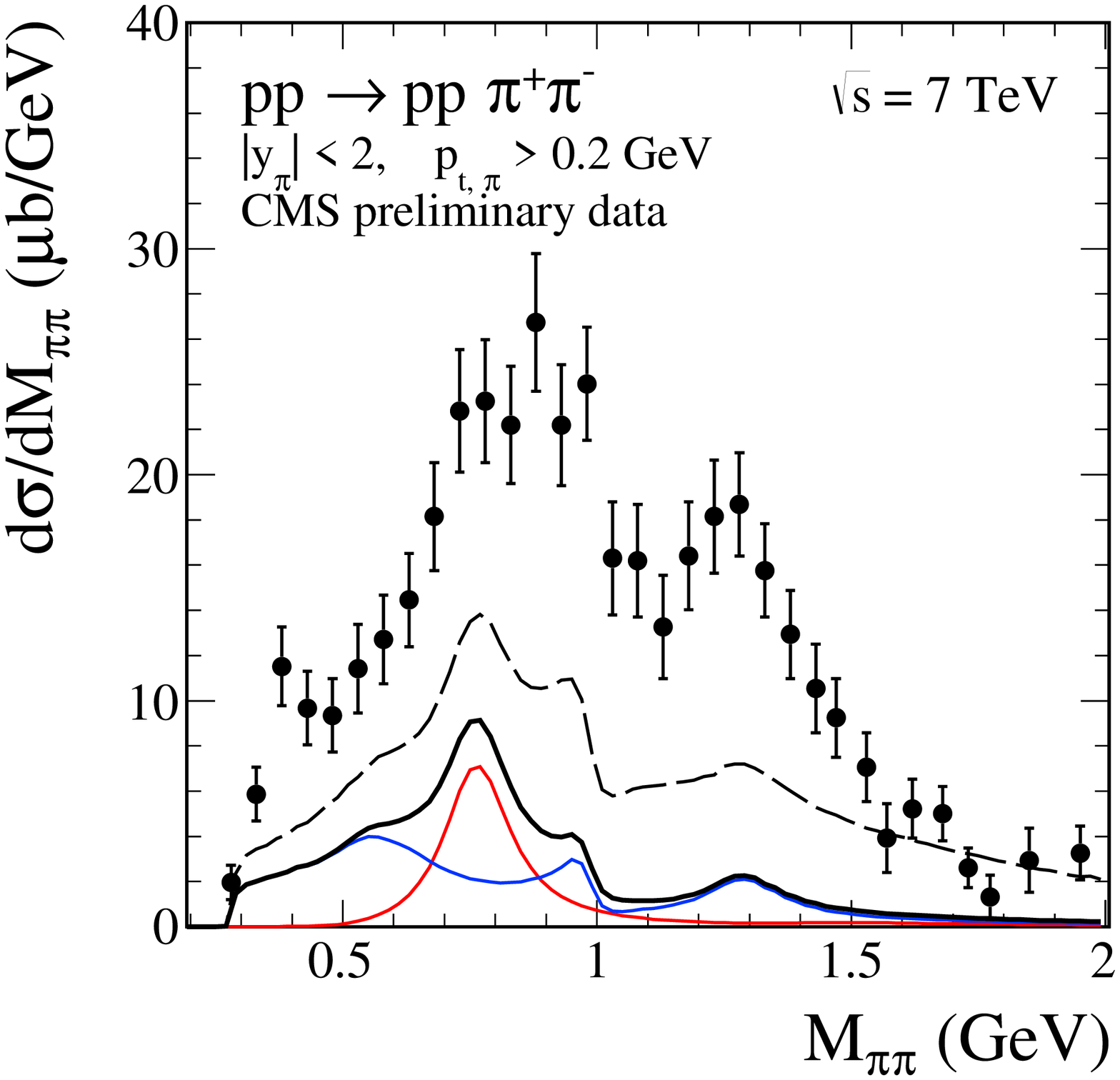}
\includegraphics[width=0.48\textwidth]{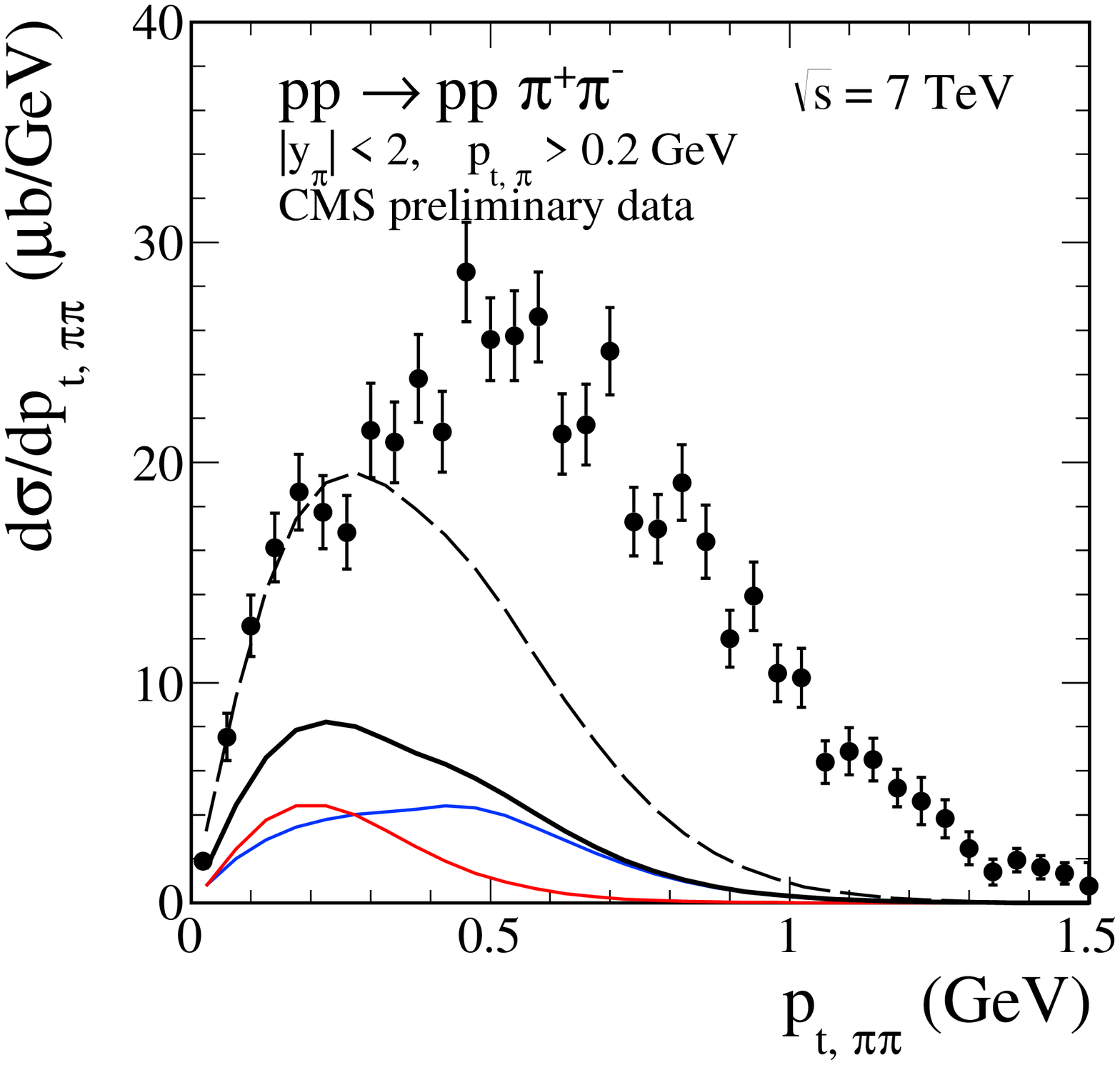}
  \caption{\label{fig:2}
  \small
The distributions for two-pion invariant mass (left panel)
and transverse momentum of the pion pair (right panel)
for the CMS kinematics at $\sqrt{s}=7$~TeV. 
Both photoproduction (red line) and purely diffractive (blue line) contributions 
multiplied by the gap survival factors $\langle S^{2}\rangle = 0.9$
and $\langle S^{2}\rangle = 0.1$, respectively, are included.
The complete results correspond to the black solid line ($\Lambda_{off,M} = 0.7$~GeV) 
and the dashed line ($\Lambda_{off,M} = 1.2$~GeV).
The CMS preliminary data \cite{CMS:2015diy} are shown for comparison.
}
\end{figure}

In Fig.~\ref{fig:2} we show results including 
in addition to the non-resonant $\pi^{+}\pi^{-}$ continuum, 
the $f_{2}(1270)$ and the $f_{0}(980)$ resonances,
the contribution from photoproduction ($\rho^{0} \to \pi^{+}\pi^{-}$, Drell-S\"oding mechanism), 
as well as the $f_{0}(500)$ resonant contribution.
Our predictions are compared with the CMS preliminary data \cite{CMS:2015diy}.
The CMS measurement \cite{CMS:2015diy} is not fully exclusive
and the $M_{\pi\pi}$ spectrum contains therefore contributions associated
with one or both protons undergoing dissociation.
Here the absorption effects lead to huge damping of the cross section 
for the purely diffractive term (the blue lines) and relatively small
reduction of the cross section for the photoproduction term (the red lines).
Therefore we expect one could observe the photoproduction term, especially at higher energies.

\section{Conclusions}

In our recent paper we have analysed the exclusive central production of
dipion continuum and resonances contributing to the $\pi^{+} \pi^{-}$ pair production
in proton-(anti)proton collisions in an effective field-theoretic approach
with tensor pomerons and reggeons.
We have included the scalar $f_{0}(500)$ and $f_{0}(980)$ resonances,
the tensor $f_{2}(1270)$ resonance
and the vector $\rho(770)$ resonance in a consistent way.
In the case of $f_{2}(1270)$-meson production via ``fusion'' of two tensor pomerons
we have found \cite{Lebiedowicz:2016ioh} all (seven) possible pomeron-pomeron-$f_{2}$ couplings
and the corresponding amplitudes using
the effective field theoretical approach proposed in \cite{Ewerz:2013kda}.
The different couplings (tensorial structures) give different results due 
to different interference effects of the $f_{2}$ resonance and the dipion continuum contributions. 
By assuming dominance of one of the couplings 
we can get only a rough description of the recent CDF and preliminary STAR experimental data.
The model parameters of the optimal coupling ($j=2$) have been roughly adjusted
to recent CDF data and then used for the predictions for the STAR, ALICE, and CMS experiments.
In the future they could be adjusted by a comparison with precise experimental data.
We have shown some differential distributions related to produced pions
as well as some observables related to final state protons,
e.g., different dependence on proton transverse momenta and 
azimuthal angle correlations between outgoing protons
could be used to separate the photoproduction term, see \cite{Lebiedowicz:2016ioh}. 

The measurement of forward/backward protons which should be possible for CMS-TOTEM and ATLAS-ALFA
is crucial in better understanding of the mechanism
of the $p p \to p p \pi^+ \pi^-$ reaction. 
The absorption effects due to $pp$ and $\pi p$ interactions
lead to a significant modification of the shape of the distributions in $\phi_{pp}$ and $p_{t,p}$
and could also be tested by these experimental groups.
Future experimental data on exclusive meson production at high
energies should thus provide good information on the spin structure of 
the pomeron and on its couplings to the nucleon and the mesons.

This research was partially supported by 
the MNiSW Grant IP2014~025173 (Iuventus Plus)
and the Polish NCN Grants DEC-2014/15/B/ST2/02528 and DEC-2015/17/D/ST2/03530.

\end{document}